# A Novel Ramp Metering Approach Based on Machine Learning and Historical Data

Anahita Sanandaji  
Ohio University

Saeed Ghanbartehrani  
Ohio University

Zahra Mokhtari  
Oregon State University

Kimia Tajik  
Oregon State University

## ABSTRACT
The random nature of traffic conditions on freeways can cause excessive congestions and irregularities in the traffic flow. Ramp metering is a proven effective method to maintain freeway efficiency under various traffic conditions. Creating a reliable and practical ramp metering algorithm that considers both critical traffic measures and historical data is still a challenging problem. In this study we use machine learning approaches to develop a novel real time prediction model for ramp metering. We evaluate the potentials of our approach in providing promising results by comparing it with a baseline traffic-responsive ramp metering algorithm.

## 1. RESEARCH PROBLEM
Ramp meter is a traffic light installed on a ramp to control traffic flow entering a freeway. Ramp metering is an effective way to reduce traffic congestion and maintain capacity flow on a freeway. Ramp metering regulates the access of ramp traffic to the mainline [1]. Figure 1. shows a ramp metering system which allows on-ramp vehicles to enter freeway only if ramp signal is green. Therefore, the vehicles need to wait behind the stop line for a green light to enter the freeway.

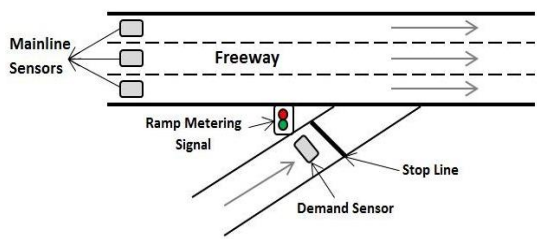

**Figure 1. Ramp metering system**

One of the most important impacts of ramp metering is to prevent freeway breakdown phenomenon. This phenomenon often occurs at freeway entrance ramps where platoons of vehicles entering the congested freeway create a bottleneck and reduce service capacity. The shock wave created by a sudden drop in speed may propagate many miles upstream causing hazardous situations. Ramp metering can minimize the shock wave effect and prevent freeway breakdown. In addition, it improves safety and reduces travel time and environmental pollution [2].

Ramp metering methods are classified into two primary categories: fixed-time control and traffic-flow responsive control. Fixed time ramp metering methods consider historical traffic information to determine the metering rates and establish the rates on a time-of-day basis [3]. Therefore, these systems do not perform effectively in presence of severe traffic fluctuations.

Traffic-responsive approaches benefit from data collected by sensors installed on freeways to calculate metering rates [2]. Local traffic responsive methods control traffic conditions on freeway by considering online traffic measures (e.g. on-ramp demand, occupancy, etc.) in the vicinity of the ramp. ALINEA [1] is an example of a promising traffic-responsive algorithm which is currently used in some ramp meter systems to reduce traffic congestion and maintain the desired freeway occupancy [2]. The inherent dynamic nature of the traffic flow makes traffic control a challenging problem that still requires more research [4].

Most of the conventional traffic-responsive ramp metering models control the ramp signal only based on the current traffic volume. In this paper, we use machine learning to develop a smart ramp metering algorithm to consider important traffic measures such as on-ramp demand, breakdown capacity [2], occupancy, and speed as well as the historical traffic data. We compare the performance of the proposed algorithm with ALINEA ramp metering method as a baseline using real traffic data available to the public.

## 2. BACKGROUND
A taxonomy of conventional ramp metering algorithms based on a study by Papageorgiou and Kotsialos [5] is presented in Figure 2. Fixed time metering is the oldest and simplest off-line strategy which is usually adjusted based on historical data and applied during particular times of day. Obviously, fixed time metering has very limited application due to its major shortcoming of not reacting to any real time traffic metrics.

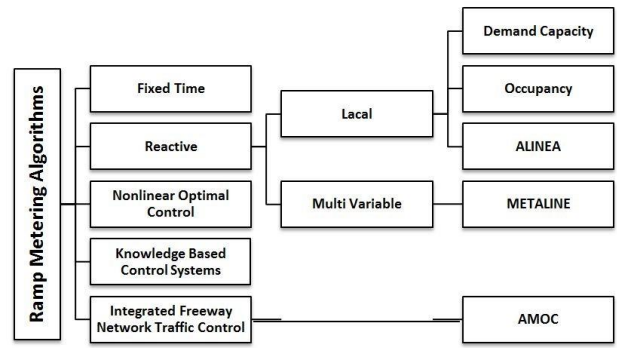

**Figure 2. Ramp metering algorithms classification**

In contrast to fixed time methods, reactive ramp metering techniques are based on real time traffic metrics. Local ramp metering uses traffic measures collected form the ramp vicinity. Demand, capacity, and occupancy based strategies allow as much traffic inflow as possible to reach the freeway capacity. ALINEA offers a more complex and more responsive strategy that unlike capacity and occupancy strategies generates smoother responses towards changes in metrics.

Multivariable regulator strategies perform the same as local strategies, but more comprehensively and independently on a set of ramps and usually outperform local strategies. METALINE can be viewed as a more general and extended form of ALINEA [6].

Reactive ramp metering strategies are helpful to some extent if their target parameters (capacity, occupancy, etc.) are set to appropriate values. They also have a local nature which may cause some control inconsistencies in a larger scale.

Nonlinear optimal control strategy considers local traffic parameters and metrics as well as nonlinear traffic flow dynamics, incidents, and demand predictions in a freeway network and outputs a consistent control strategy. This introduces many challenges including quite high computational load and complexity in design and application of corresponding models.

Knowledge based control systems are developed based on historical data and human expertise. They usually rely on heuristics and ad-hoc procedures for traffic control to provide a good and not necessarily optimal output. Integrated freeway network traffic control is a more general approach to nonlinear control that extends application of optimal control strategies to all forms of freeway traffic control.

All mentioned strategies have their own strengths and weaknesses, but in general, more complex models were introduced to address simpler models' shortcomings. A common drawback of model based controls is their sensitivity to the model's accuracy. Practically, mathematical traffic models can seldom represent the full, real world traffic dynamics [6]. Also, models rely on accurate parameters and full information regarding the system they represent, which is not always easy to achieve. On the other hands, almost all approaches that apply control theory are computationally demanding and therefore impractical for real world applications [7].

In case of knowledge based systems, inability to learn and adapt to temporal evolution of the system being controlled can be an issue, so knowledge based systems need to be periodically updated to remain efficient [7].

All these issues together, urge the quest to develop an intelligent algorithm which can be adapted to different settings and scenarios at a reasonable cost, without the need of very accurate parameters and enormous processing infrastructure. Artificial intelligence and machine learning approaches seem to be eligible candidates for this purpose. Several intelligent methods have been proposed for ramp metering including reinforcement learning (RL) [8] and artificial neural networks (ANN) [9]. Research presented in [4], [7] and [10]–[14] describes some of the models that use reinforcement learning. Most of the reinforcement learning models use Q-learning [15] or some methods based on Q-learning. Models in [16]–[18] are based on ANN. All these models are trained and benchmarked using computer simulation and none of them were implemented in a real ramp meter. This fact can raise doubt over practicality and efficiency of the models mentioned in comparison to the ones already in use.

For most of machine learning techniques, training phase is very critical and final performance and efficiency of the algorithm strongly depends on that. In some cases, researchers tried to introduce incidents and accidents to gain a well prepared model. However, there can still be many unpredictable situations for which the proposed model would not be trained for. A possible solution we are proposing is to use real historical data captured from different ramp locations to train the models instead of or along with the use of simulation models. Additionally, we focus on integrating regression and clustering, two effective machine learning approaches, to come up with a framework to control ramp signal in an accurate and efficient way.

## 3. METHODOLOGY

The proposed ramp metering algorithm is shown in Figure 3. Our method consists of four main modules: 1) data refinement and feature selection; 2) regression; 3) clustering; and 4) ramp metering algorithm. In the following, we will discuss our proposed approach for each module.

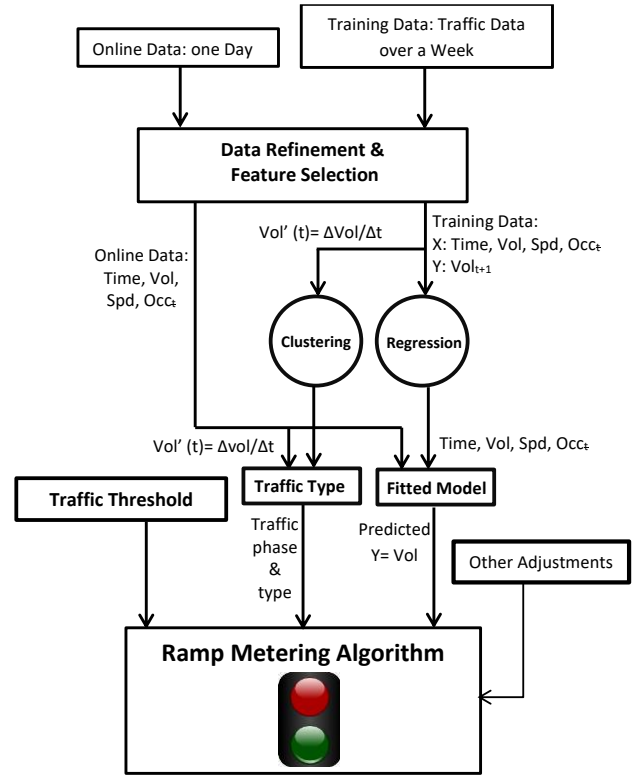

**Figure 3. The general schema of proposed algorithm**

### 3.1 Data Refinement and Feature Selection

We chose a freeway section of I-205, in the state of Oregon, and analyzed real traffic data collected at this station for a working week [19]. As shown in Figure 3., the inputs to our algorithm are: 1) the traffic data captured over the working days of one week to train the algorithm; and 2) the online traffic data for a one day period to test the algorithm. This data is passed to the data refinement and feature selection module. In this module, we selected time, occupancy, volume, and speed data features.

### 3.2 Regression

In the regression module, we used linear regression to predict *Vol(t+1)* based on *Time(t)*, *Occupancy(t)*, *Speed(t)*, and *Vol(t)* data. To validate the regression results, we compared predicted data to real data and observed full conformance. The regression model was perfectly consistent with the trend observed in real data as shown on Figure 4.

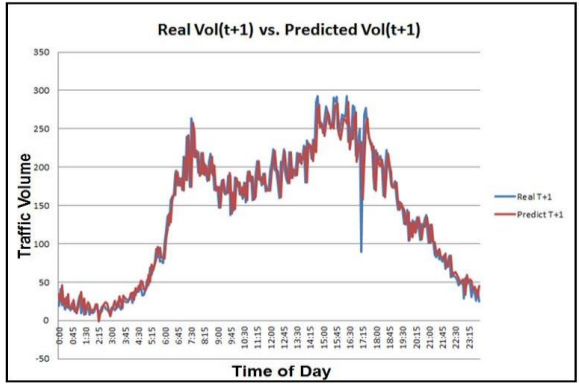

**Figure 4. The estimated volume matches the real one perfectly**

## 3.3 Clustering

In this part, we implement two k-means clustering approaches. The first set of clusters is to determine traffic phases in which we cluster data based on *Time* and $\Delta vol/\Delta t$ to identify traffic phases capturing the fluctuations in traffic volume over time. We identified five traffic phases (1-5) represented with the five clusters in Figure 5.

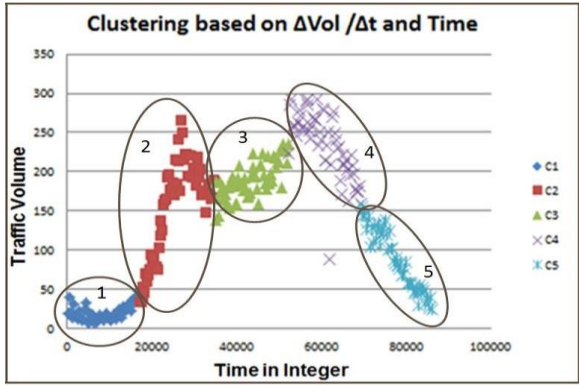

**Figure 5. The five identified traffic phases**

In the second set of clusters, we cluster data only based on $\Delta vol/\Delta t$ to determine traffic type. This cluster is used for capturing the fluctuations in traffic volume.

**Table 1. Two cluster sets and their definitions**

| K | Cluster 1: Traffic Phase | Cluster 2: Traffic Type |
|---|---|---|
| 1 | 0:00 – 4:40 AM (early morning) | Sharp negative slope : -- |
| 2 | 4:45-9:35 AM (morning) | Moderate negative slope : - |
| 3 | 9:40AM-2:25 PM ( afternoon) | Small slope or constant almost straight line : 0 |
| 4 | 2:30 -7:10 PM ( evening) | Moderate positive slope : + |
| 5 | 7:15 -23:55 (night) | Sharp positive slope : ++ |

The five traffic phases identified in the first set of clusters are presented in the second column of Table 1. The third column contains the second set, reflecting the five traffic types identified based only on $\Delta vol/\Delta t$. Figure 6. shows traffic volume data points colored based on their corresponding traffic type clusters.

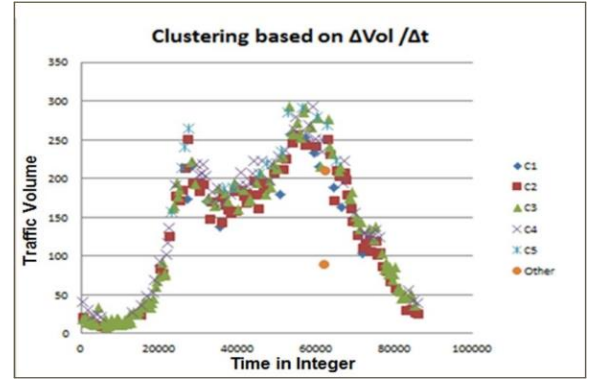

**Figure 6. The five traffic types identified based on $\Delta vol/\Delta t$.**

## 3.4 Proposed Ramp Metering Algorithm

The schematic view of our final ramp metering algorithm (RMA) is shown in Figure 7. Results of the regression and clustering modules are fed to the RMA.

The RMA starts by first computing the prediction error $PE(t)$ which is the discrepancy between the estimated traffic volume $V_{Est}(t)$ found by the regression module and the real traffic volume $V_{Real}(t)$ based on the online traffic data.

$$PE(t) = V_{Real}(t) - V_{Est}(t)$$

The goal is to make sure that the prediction is within an acceptable threshold. We computed the threshold based on the negative and positive average errors of real traffic data over a week period and fine tuned it with some try and error. If estimation error is not within the given thresholds, the corrected $V_{Est}(t+1)$ is calculated as follows:

$$V_{Est}(t + 1) = V_{Est}(t + 1) \times (1 - PE(t))$$

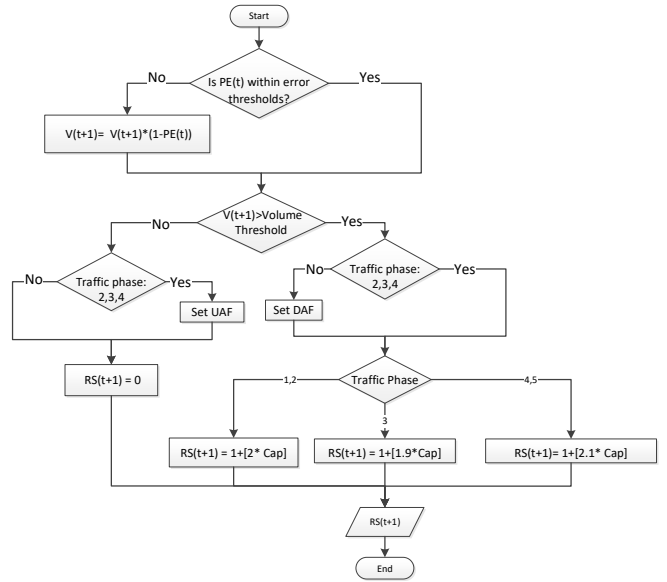

**Figure 7. Proposed final ramp meter algorithm schematic**

Next, we compare estimated volume (or the corrected estimated volume) using volume threshold. We calculated the volume threshold for five minutes (size of each time slot), considering average length of vehicles (15 ft., which is found by averaging the length of all vehicles traveling in the same section of the studied highway for a few months), maximum allowed speed limit of the highway (55 mph), and the safe distance of two vehicles (2 seconds or 161 ft.) [21].

The ramp signal status is decided based on the current traffic phase determined by the clustering module and the result of the comparison performed in the previous step. The ramp signal is set to *S0* (disabled) if the estimated volume is less than the threshold. In case of one of the 2, 3, or 4 traffic phases, we raise *UAF* (possible Upstream Accident Flag) since traffic volumes less than the threshold are not expected at these phases and can be caused by an upstream accident. If the estimated volume is greater than threshold while in traffic phases 1 or 5, we set *DAF* (possible Downstream Accident Flag) since high traffic volume at these phases are not expected and can be caused by an accident downstream.

At the end, the next state of the ramp signal *RS(t+1)* is computed based on the traffic type (determined by the clustering module) and capacity which is calculated as follows:

$$Cap = V_{Est}(t+1)/250$$

The next state of the ramp signal *RS(t+1)* is:

$$RS(t+1) = 1 + [\alpha \times Cap]$$

Brackets [.] represent the floor function and the constant $1.9 \leq \alpha \leq 2.1$ is determined based on the desired ramp signal policy. In this instance, we set $\alpha$ to values closer to (2) in case of traffic types 1, 2, 4 or 5 (i.e. highly increasing or decreasing volume) which translates into more aggressive ramp metering (i.e. signals with higher delay). $\alpha$ is set to lower values (1.9) in case of traffic type 3 (steady traffic volume) which results in less aggressive (i.e. signals with shorter delays) ramp metering.

Table 2. presents ramp signal values and their interpretation.

**TABLE 2. Ramp signals**

| Signal | Description |
|---|---|
| S0 | Ramp signal is disabled |
| S1 | Ramp signal is active with short delay |
| S2 | Ramp signal is active with medium delay |
| S3 | Ramp signal is active with long delay |

## 4. EVALUATION AND RESULTS

In this section, we compare the proposed algorithm with the widely used traffic-responsive algorithm, ALINEA using the same dataset.

ALINEA employs closed-loop feedback for green-phase duration as follows:

$$g(k) = g(k-1) + K_R \frac{C}{r_{sat}}(\hat{O} - O_{out}(k)), g_{min} \leq g \leq g_{max}$$

$g(k)$: Green-phase duration at time interval *k* (in seconds)

$g(k-1)$: Green-phase duration at time interval *k-1* (in seconds)

$C$: The fixed signal cycle duration (red phase + green phase) (in seconds)

$r_{sat}$: The ramp capacity flow (vehicles/hour)

$K_R$: Regulator parameter (vehicles/hour)

$\hat{O}$: Critical occupancy (%)

$O_{out}(k)$: Occupancy downstream of the merge area at time interval *k* (%).

In this study, we set $K_R = 70$, $C = 8$ *sec.*, $r_{sat} = 250$ *vehicles/hour*, and $\hat{O} = 3\%$ for regular hours and $\hat{O} = 9\%$ for rush hours. The values for $r_{sat}$ and $\hat{O}$ are calculated based on the test data. Moreover, $K_R = 70$ is the recommended value by the majority of the analytical studies that used ALINEA. Table III presents ALINEA's ramp metering policy.

Table 3. presents ALINEA ramp metering policy.

**TABLE 3. ALINEA ramp signal policy**

| Green-phase duration | Ramp signal |
|---|---|
| green − phase duration = 0 | S3: Ramp signal is active with long delay |
| 0 < green − phase duration ≤ 1 | S2: Ramp signal is active with medium delay |
| 1 < green − phase duration ≤ 5 | S1: Ramp signal is active with short delay |
| 5 < green − phase duration ≤ 8 | S0: Ramp signal is disabled |

The comparison results are presented in Figure 8.

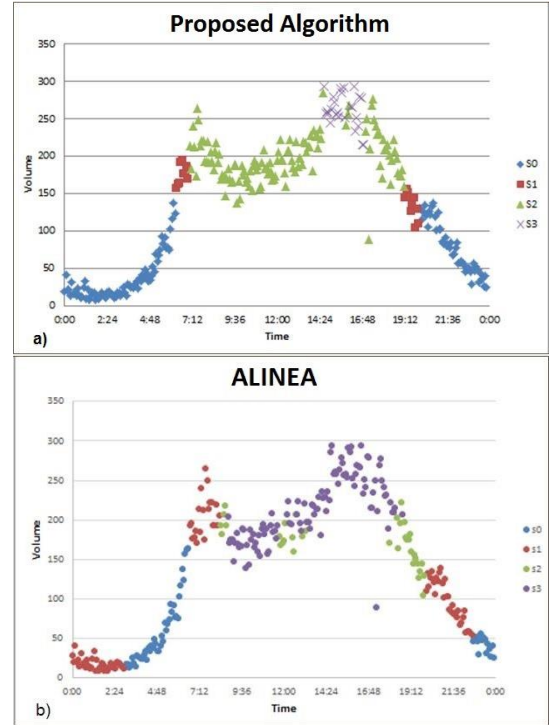

**Figure 8: (a) proposed algorithm vs (b) ALINEA**

The proposed algorithm shows reasonable ramp signal control and performs more permissively in allowing vehicles to enter the freeway compared to ALINEA. This means that the proposed algorithm offers lower delays compared to ALINEA as long as the critical capacity is not reached. This behavior is also noticeable during rush hours when the proposed algorithm outputs medium delay while ALINEA strictly generates long delays.

Pie charts in Figure 9. illustrate the distribution of different generated ramp signals for each algorithm. The higher frequency of S0 and S2 (blue and green slices) compared to S1 and S3 (red and purple slices) generated by the proposed algorithm indicates its tendency towards generating shorter delays compared to ALINEA which tends to behave more strictly even in situations that there is no need for long delays in the ramp signal.

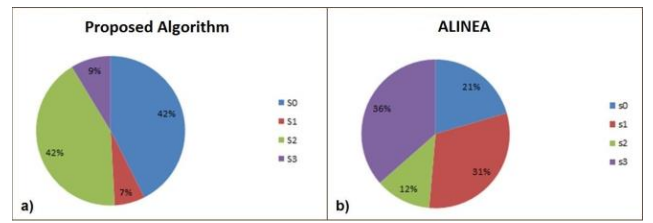

**Figure 9. (a) Proposed algorithm vs (b) ALINEA signals**

Figure 10. compares the performance of both algorithms during traffic phases 1 and 5 (i.e. off-peak phases, early mornings and late nights). Despite phase 1 and 5's inherent low traffic flow, ALINEA still generates excessively long delays (S3) very frequently compared to the proposed algorithm.

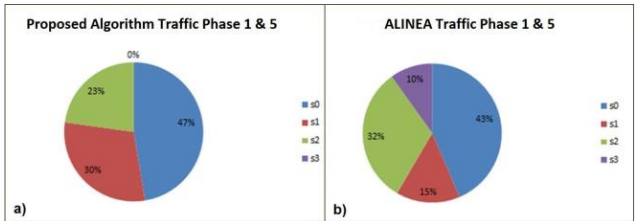

**Figure 10. Comparison of signals in Phase 1 and 5**

A similar trend is visible in Figure 11. which shows the distribution of signals in peak phases 2, 3, and 4.

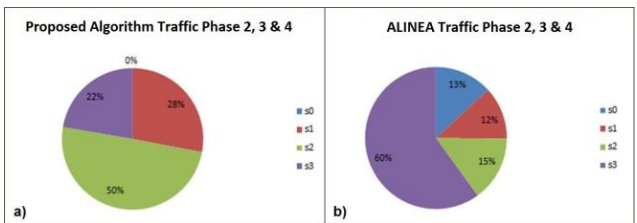

**Figure. 11: Comparison of signals in Phase 2, 3, and 4**

The proposed algorithm is capable of maintaining the freeway flow below the critical capacity mostly by medium delays (S2, green slice) while ALINEA resorts to long delays (S3, purple) more frequently.

## 5. CONCLUSIONS

This research investigated the potentials of using machine learning techniques to develop a ramp signal control model. The presented model incorporates linear regression and clustering approaches to learn the traffic flow trend over time. We compared the proposed algorithm with the widely used traffic-responsive algorithm, ALINEA using the same data.

The results of the comparison confirms that the proposed algorithm can effectively maintain freeway traffic flow at reasonable levels while allowing the on ramp traffic into the mainline as much as possible. However, ALIENA performed more conservatively by limiting the traffic flow even during traffic phases with minimal traffic flow.

We can achieve a more in depth analysis and comparison on the performance of the proposed algorithm by a simulation study as an area for future work. In addition, by running a simulation study, we can evaluate the irregularities in traffic flow detected by the proposed algorithm (i.e. accident flags). The results of the simulation study can be validated using real traffic flow data.